# UNIVERSAL ASYMPTOTIC UMBRELLA
# FOR HYDRAULIC FRACTURE MODELING

### Alexander M. Linkov

*Institute for Problems of Mechanical Engineering, 61, Bol'shoy pr. V. O., Saint Petersburg, 199178, Russia*
*Rzeszow University of Technology, ul. Powstancow Warszawy 8, Rzeszow, 35-959, Poland*
*ve-mail: linkoval@prz.edu.pl*

**Abstract.** The paper presents universal asymptotic solution needed for efficient modeling of hydraulic fractures. We show that when neglecting the lag, there is universal asymptotic equation for the near-front opening. It appears that apart from the mechanical properties of fluid and rock, the asymptotic opening depends merely on the local speed of fracture propagation. This implies that, on one hand, the global problem is ill-posed, when trying to solve it as a boundary value problem under a fixed position of the front. On the other hand, when properly used, the universal asymptotics drastically facilitates solving hydraulic fracture problems (both analytically and numerically). We derive simple universal asymptotics and comment on their employment for efficient numerical simulation of hydraulic fractures, in particular, by well-established Level Set and Fast Marching Methods.

## 1. INTRODUCTION

Hydraulic fracturing is a technique widely used in many engineering applications; it also occurs as a natural process in the earth crust. Because of difficulties of obtaining field data on the hydraulic fracture (HF) propagation, much theoretical work has been done to obtain guidelines for engineering practice (see, e.g. reviews in the papers by Adachi et al. (2007) and in many papers in the Proceedings of the International Conference "Effective and Sustainable Hydraulic Fracturing" (2013). Since the papers by Nordgren (1972), Spence & Sharp (1985) and Kemp (1989), the importance of proper accounting for the asymptotics near the fracture contour has been comprehended, studied and used for obtaining analytical and numerical solutions (e.g. Desroches et al. 1994; Lenoach 1995; Garagash & Detournay 2000; Savitski & Detournay 2002; Garagash 2006; Mitchell et al. 2007; Kovalyshen & Detournay 2009; Garagash et al. 2011). The importance of their using has been recently emphasized and demonstrated in the papers by Linkov (2012), Mishuris et al. (2012), Linkov & Mishuris (2013); it is also of prime significance for the implicit level set algorithm (ILSA), suggested by Peirce and Detournay (2008), which efficiency has been confirmed in the recent papers by Gordeliy & Peirce (2013) and Lecampion et al. (2013).

When analyzing the Nordgren problem, it has been also revealed that the asymptotics of the opening is completely defined by the speed of the fracture propagation. Then the speed equation (Kemp 1989; Linkov 2011, 2012) is satisfied identically. This makes the problem ill-posed when trying to solve it as a boundary value problem for a fixed position of the front. The same is also the case in other cases (Linkov 2012). Meanwhile, when properly used, the SE written with employing the asymptotics, suggests notable analytical and computational gains (Linkov 2012; Mishuris et al. 2012; Linkov& Mishuris 2013; Linkov 2013, 2014).

The paper aims to make a further step in understanding and solving the HF problem by obtaining universal dependences which facilitate developing efficient numerical algorithms. In *Sec. 2*, we show that, when neglecting the lag, the near-front equations present a universal system, in which the only characteristic, incorporating features of a particular global motion, is the local fracture propagation speed $v_*$. Therefore, for zero lag, there exists the universal asymptotic solution depending merely on the local propagation speed. This leads to important analytical and computational consequences discussed in *Sec. 3*. We note that, on one hand, the global HF problem appears ill-posed when neglecting the lag and trying to solve it as a boundary value problem with a fixed position of the front on a time step. Still, it may be efficiently solved by using appropriate regularization technique, in particular, $\varepsilon$-regularization. On the other hand, the universal asymptotics, when properly used, provides a means for developing highly efficient numerical schemes to model HF. They are based on including the speed equation into the system of equations with the front position $\mathbf{x}_*$ as an additional unknown. Since any of the approaches is markedly facilitated by using the universal asymptotic solution, we present analytical equations, defining the universal asymptotic opening for power-law fluids and power-law leak-off. They are given in Sec. 3 in simple monomial forms for toughness, viscosity and leak-off dominated regimes of fracture propagation and for regimes intermediate between these limiting cases. *Sec. 4* contains brief comments on the ways to increase the efficiency of numerical modeling by using the universal asymptotics.

## 2. NEAR-FRONT EQUATIONS

*2.1. Near-front elasticity equations.* For a narrow strip near the fracture contour, the asymptotic form of the hypersingular integral equations of 3D elasticity for a region with a crack (e.g. Ioakimidis 1982a,b,1987; Linkov & Mogilevskaya 1986; Linkov 1995; Peirce & Napier 1995) are reduced to the *elasticity* plane-strain *equation* (e .g. Peirce and Detournay 2008):

$$\frac{E'}{4\pi} \int_0^\infty \frac{\partial w}{\partial y} \frac{dy}{r-y} = p(r) \,, \tag{2.1}$$

where $E' = \frac{E}{1-v^2}$, $E$ is the Young's modulus, $v$ is the Poisson's ratio, $w$ is the crack opening, $r$ is the distance from the crack tip; in HF problems, at the scale of asymptotic analysis, $p$ is the net-pressure, that is the difference between the fluid pressure and the far-field traction normal to the crack surface. Equation (2.1) corresponds to a problem of plain-strain elasticity for a plane with a semi-infinite crack ($0 \le r < \infty$) with the normal component of the displacement discontinuity $w(r)$ and the normal component of traction $p(r)$. It is solved under the condition that the opening is zero at the crack tip:

$$w(0) = 0. \tag{2.2}$$

*2.2. Near-front fluid equations.* For points near the front, the *continuity equation* (CE) is (see, e. g. Peirce and Detournay 2008; Linkov 2012):

$$\frac{\partial w}{\partial t} + \frac{\partial}{\partial r}[(v_* - v)w] + q_l = 0 \,, \tag{2.3}$$

where $v_*$ is the speed of the fracture propagation, $v$ is the fluid particle velocity averaged over a cross section, $q_l$ is the density of sources and/or sinks, distributed in the channel; we assume it positive in the case, when this term corresponds to fluid leak-off from a HF into a rock formation. In further discussion, it is assumed to be prescribed by the power-type dependence:

$$q_l = \frac{2C_l}{(t-t_*)^{\beta_l}}, \qquad\qquad 0 \le \beta_l < 1, \tag{2.4}$$

where $C_l$ and $\beta_l$ are constants of leak-off, $t - t_*$ is the time elapsed since the moment $t_*$, when the fluid front had reached a considered point. In the commonly considered case of Carter's leak-off (Carter, 1957), $\beta_l = 1/2$. Near a tip moving with the speed $v_*$, the leak-off term (2.4) becomes:

$$q_l = \frac{2C_l v_*^{\beta_l}}{r^{\beta_l}}, \qquad\qquad 0 \le \beta_l < 1. \tag{2.5}$$

From now on, we shall focus on the case most important for applications when the *lag is negligibly small*. Then the fluid front coincides with the fracture contour. In view of the boundary condition (2.2) and common power-type asymptotics of the opening, the convective term in (2.3) dominates the time derivative. Neglecting the time derivative and integrating in $r$, the CE (2.3) reads:

$$v = v_* + \frac{1}{w}\int_0^r q_l\,dr \,. \tag{2.6}$$

Equation (2.6) corresponds to steady propagation of the fracture tip with the speed $v_*$. For the power-law leak-off term (2.5), it becomes:

$$v = v_* + \frac{2C_l v_*^{\beta_l} r^{1-\beta_l}}{(1-\beta_l)w} \,. \tag{2.7}$$

The CE (2.13) is derived under the assumption that the flow occurs predominantly in the fracture plane. Hence, this assumption refers to (2.7), as well. This implies that strictly speaking leak-off should be either finite at the crack tip ($\beta_l = 0$), or its singularity less than that of $\frac{\partial w}{\partial r}$. Otherwise, the assumption that the flow occurs predominantly in the fracture plane is violated. In particular, this happens when using the Carter's value $\beta_l = 1/2$ for leak-off at the fracture tip. Indeed, in this case, the second term on the r. h. s. in (2.7) goes to infinity as $r \to 0$. Since leak-off occurs in the directions normal to the channel walls, this means that the component of the particle velocity normal to the wall is much greater than the in-plane component, what violates the accepted assumptions.

In the physically consistent case, the second term on the r. h. s. of (2.7) goes to zero when $r \to 0$. Then in the limit, the CE leads to the speed equation (SE) (Kemp 1989; Linkov 2011, 2012):

$$\lim_{r\to 0} v = v_* \,. \tag{2.8}$$

We see that in the case of zero lag and physically consistent leak-off, any solution of the CE, satisfying the boundary condition of zero-opening (2.2), automatically meets the SE (2.8). This non-trivial fact yields far-reaching consequences for the problem under consideration (Linkov 2011, 2012).

Still, having in mind that the Carter's dependence for the leak-off term is commonly used for the HF problem, we shall not exclude the case when the second term on the r. h. s. of (3.3) goes to infinity. It is especially important for numerical simulation of HF, when Carter's leak-off may present a reasonable approximation at a zone behind the lag.

The CE is complimented by the *equation of motion* of a fluid in a narrow channel. For HF problems, a fluid may be assumed incompressible (see, e.g. Spence & Sharp 1985). Its viscous resistance is usually described by the power-law:

$$\sigma_{z\tau} = M\dot{\gamma}^n, \tag{2.9}$$

where $\sigma_{z\tau}$ is the shear stress on that wall, for which the axis $z$ is directed inside the channel, $\tau$ is the direction of the particle velocity, $\dot{\gamma} = 2\dot{\varepsilon}_{z\tau}$ is the doubled shear strain rate, $M$ is the consistency index, $n$ is the behavior index. Using (2.9) in the common derivation of the Poiseuille type dependence for a flow in a narrow channel yields:

$$v = \left[\frac{w^{n+1}}{\mu'}\left(-\frac{\partial p}{\partial \tau}\right)\right]^{1/n}, \tag{2.10}$$

where

$$\mu' = 2\left(2\frac{2n+1}{n}\right)^n M. \tag{2.11}$$

After substitution (2.10) into (2.7) and noting that $\tau = -r$, we obtain the lubrication equation for the near-tip region:

$$\left[\frac{w^{n+1}}{\mu'}\frac{\partial p}{\partial r}\right]^{1/n} = v_* + \frac{2C_l v_*^{\beta_l} r^{1-\beta_l}}{(1-\beta_l)w}. \tag{2.12}$$

In the limit $r \to 0$, the lubrication equation (2.12) yields:

$$\lim_{r\to 0}\left[\frac{w^{n+1}}{\mu'}\frac{\partial p}{\partial r}\right]^{1/n} = v_* + \lim_{r\to 0}\frac{2C_l v_*^{\beta_l} r^{1-\beta_l}}{(1-\beta_l)w} \tag{2.13}$$

When the leak-off is neglected or small, equation (2.13) turns into the standard SE (2.8). Equation (2.13) may be called *the generalized SE for strongly singular leak-off*. Emphasize, that in accordance with the physical essence of the problem, *the propagation speed $v_*$ is always assumed finite and non-zero*. This is a key assumption accepted, often implicitly, in all papers on HF. It is basic for studying asymptotic behavior of the solution.

*2.3. Fracture propagation condition.* The fracture propagates when the strength of rock is reached. In the case of the tensile mode of fracture, the condition of linear fracture mechanics is (see, e.g. Rice 1968):

$$K_I = K_{Ic}, \tag{2.14}$$

where $K_I$ is the normal stress intensity factor (SIF) at a given point of the contour, $K_{Ic}$ is its critical value defined by the strength of rock. The elasticity theory yields, that the condition (2.14) may be written in terms of the limit of the opening, satisfying the condition (2.2), as (e.g. Rice 1968):

$$\lim_{r\to 0}\frac{w(r)}{r^{1/2}} = \sqrt{\frac{2}{\pi}}\frac{4}{E'}K_{Ic}. \tag{2.15}$$

Equation (2.15) is also true when the fracture toughness may be neglected ($K_{Ic} = 0$).

*2.4. Universal equations.* Summarizing, we have the elasticity equation (2.1), the lubrication equation (2.12) and the limit equation at the crack tip (2.15). They correspond to steady propagation of a semi-infinite crack ($0 \le r < \infty$) with the speed $v_*$ when the lag is neglected and the mode of fracture is tensile with the critical SIF $K_{Ic}$. Therefore we have the system of two equations (2.1), (2.12) in two unknowns $w$ and $p$ with the condition (2.15), defining the asymptotic behavior of the solution at the tip $r = 0$.

Note that equations (2.1), (2.12) and (2.15) are *universal* in the sense that they do not include characteristics of the fracture surface, fluid motion and fracture propagation outside a vicinity of a considered point on the fracture contour. The only characteristic, which implicitly incorporates features of a particular global motion, is the speed $v_*$ of the fracture propagation at the point.

# 3. UNIVERSAL ASYMPTOTIC UMBRELLA

*3.1. Universal asymptotic solution. Analytical and computational consequences.* In the previous section, we have come to the universal system on the half-axis $r \ge 0$.:

$$\frac{E'}{4\pi}\int_0^\infty \frac{\partial w}{\partial y}\frac{dy}{r-y} = p(r), \tag{3.1}$$

$$\left[\frac{w^{n+1}}{\mu'}\frac{\partial p}{\partial r}\right]^{1/n} = v_* + \frac{2C_l v_*^{\beta_l} r^{1-\beta_l}}{(1-\beta_l)w}, \tag{3.2}$$

$$\lim_{r\to 0}\frac{w(r)}{r^{1/2}} = \sqrt{\frac{2}{\pi}}\frac{4}{E'}K_{Ic}. \tag{3.3}$$

The system does not involve the opening and the net-pressure at the fracture outside a vicinity of a considered point. Therefore, *in the near-tip region, the opening w (and consequently the net-pressure p) depends on the local fracture propagation speed $v_*$ only*:

$$w = \varphi(r, v_*). \tag{3.4}$$

*Conclusion:* when neglecting the lag, the HF problem has the universal asymptotic solution at each point of the front, depending merely on the local propagation speed.

The conclusion has important consequences, both analytical and computational. Indeed, the universal solution (3.4) implies that at each point $\boldsymbol{x}_*$ of the front, in addition to the boundary condition $w(\boldsymbol{x}_*) = 0$ of zero opening we have also prescribed the normal derivative of the opening $\frac{\partial w}{\partial r} = \frac{\partial}{\partial r} \varphi(r, v_*)_{r=0}$. Thus for the elliptic in spatial coordinates global system, we have actually prescribed both the function and its normal derivative. Hence, for a fixed position of the front and fixed propagation speed, the global problem appears ill-posed in the Hadamard sense (Hadamard1902) (see also Lavrent'ev & Savel'ev 1999). This indicates that there might be difficulties if trying to solve the HF problem as a BV problem at a time step, say, by applying the Crank-Nicolson scheme. In such cases, one needs to use regularization techniques (Tychonoff 1963).

The fact that the HF problem is ill-posed when fixing the front position has been disclosed for the Nordgren formulation of the problem (Linkov 2011). The appropriate highly efficient $\varepsilon$-regularization has been suggested and successfully employed (Linkov 2011; Mishuris et al. 2012; Linkov & Mishuris 2013).

Another way to avoid difficulties consists in complimenting the system with the SE equation (2.8), containing the front position $\boldsymbol{x}_*$ as the additional unknown. (In the case of strongly singular leak-off, the generalized form (2.13) of the SE is used.) Then, we have a well-posed dynamic system (DS), which can be efficiently solved by conventional methods like the Runge-Kutta method (Mishuris et al. 2012; Linkov & Mishuris 2013). It can be also embedded in frames of the conventional LSM and FMM (Sethian 1999). Any of the approaches is markedly facilitated by using the universal asymptotic solution (3.4) (see comments in Sec. 4). Therefore, it is reasonable to obtain its explicit forms.

*3.2. Universal system in dimensionless variables.* It is convenient to re-write the system (3.1), (3.2), (3.3) in dimensionless variables and parameters:

$$\xi = \frac{r}{L_\mu}, \ \Omega = \frac{w}{L_\mu}, \Pi = \frac{p}{E\prime}, k = \left(\frac{L_k}{L_\mu}\right)^{1/2}, \ \psi = \frac{v_l}{v_*}, L_\mu = t_n v_*, L_k = \frac{32}{\pi}\left(\frac{K_{IC}}{E\prime}\right)^2, v_l = \frac{2C_l}{(1-\beta_l)t_n^{\beta_l}}, t_n = \left(\frac{\mu\prime}{E\prime}\right)^{\frac{1}{n}}, \tag{3.5}$$

where $\mu\prime$ is defined by (2.11).

In terms of these variables, the system becomes:

$$\frac{1}{4\pi}\int_0^\infty \frac{d\Omega}{d\eta}\frac{d\eta}{\xi-\eta} = \Pi(\xi), \tag{3.6}$$

$$\left(\Omega^{n+1}\frac{d\Pi}{d\xi}\right)^{\frac{1}{n}} = 1 + \psi\frac{\xi^{1-\beta_l}}{\Omega(\xi)}, \tag{3.7}$$

$$\lim_{\xi\to 0}\frac{\Omega(\xi)}{\xi^{1/2}} = k. \tag{3.8}$$

It contains two parameters only: the leak-off parameter $\psi$ and the toughness parameter $k$. Since equation (3.7) contains also the term 1, there are three limiting cases, discussed below. The parameter $\psi$ may be excluded by renormalizing the opening, net-pressure, coordinate and the coefficient $k$ as:

$$\Omega_1 = \frac{\Omega}{\psi^{\alpha_\mu/\varepsilon_\mu}}, \; \Pi_1 = \Pi\psi^{(1-\alpha_\mu)/\varepsilon_\mu}, \; \xi_1 = \frac{\xi}{\psi^{1/\varepsilon_\mu}}, \;\; k_1 = \frac{k}{\psi^{(\alpha_\mu - 1/2)/\varepsilon_\mu}}, \tag{3.9}$$

with

$$\alpha_\mu = \frac{2}{n+2}, \;\; \varepsilon_\mu = \alpha_\mu - (1 - \beta_l). \tag{3.10}$$

Then the system becomes:

$$\frac{1}{4\pi}\int_0^\infty \frac{d\Omega_1}{d\eta_1}\frac{d\eta_1}{\xi_1 - \eta_1} = \Pi_1(\xi_1), \tag{3.11}$$

$$\left(\Omega^{n+1}\frac{d\Pi}{d\xi_1}\right)^{\frac{1}{n}} = 1 + \frac{\xi_1^{1-\beta_l}}{\Omega(\xi_1)}, \tag{3.12}$$

$$\lim_{\xi_1 \to 0}\frac{\Omega_1(\xi_1)}{\xi_1^{1/2}} = k_1. \tag{3.13}$$

Actually equations (3.11)-(3.13) are less convenient than (3.6)-(3.8), because the parameter $\varepsilon_\mu$, defined in (3.10), is quite small what drastically extends the scales of $\xi_1$ and $\Omega_1$. Besides, as clear from (3.9), they are inapplicable to the case of negligible leak-off, for which $\psi = 0$. For these reasons, we shall employ them only for intermediate calculations.

*3.3. Limiting regimes*. From (3.6)-(3.8), it follows that there are three limiting cases, which correspond to three well-known (see, e.g. Garagash et al. 2011) limiting regimes of flow. In all the three cases, the exact solution is given by the monomial equation, which identically satisfies the elasticity equation (3.6):

$$\Omega(\xi) = A\xi^\alpha, \;\; \Pi(\xi) = -AB(\alpha)\xi^{\alpha - 1}, \tag{3.14}$$

where

$$B(\alpha) = \frac{\alpha}{4}\cot[\pi(1-\alpha)]. \tag{3.15}$$

The exponent $\alpha$ and the coefficient $A$ depend on a dominating factor of fracture propagation.

(i) *Toughness dominated regime* occurs when the asymptotics of the opening is defined by equation (3.3) only. Hence, in (3.14):

$$\alpha = \alpha_k = \frac{1}{2}, A = A_k = k, \tag{3.16}$$

and the value $L_\mu = 1$ is used in the normalizing quantities (3.5).

(ii) V*iscosity dominated regime* occurs when toughness and leak-off are negligible ($k = 0$, $\psi = 0$). Then the asymptotics of the opening is defined by equations (3.6) and (3.7) with $\psi = 0$. In this case,

$$\alpha = \alpha_\mu = \frac{2}{n+2}, A = A_\mu = [(1-\alpha)B(\alpha)]^{-\frac{1}{n+2}}. \tag{3.17}$$

For a Newtonian fluid ($n = 0$), equations (3.17) give the Spence and Sharp (1985) exponent $\alpha = \frac{2}{3}$, while $A_\mu = 2^{1/3}3^{5/6} = 3.147345$.

(iii) *Leak-off dominated regime* occurs when toughness is negligible ($k = 0$) while the second term on the r. h. s. of (3.7) is much greater than 1. Then insertion of (3.14) into (3.7) yields:

$$\alpha = \alpha_l = \frac{n(1-\beta_l)+2}{2n+2}, \, A = A_l = [(1-\alpha)B(\alpha)\psi^n]^{-\frac{1}{2n+2}}. \tag{3.18}$$

For Carter's leak-off ($\beta_l = 1/2$), equations (3.18) give the solution by Lenoach (1985); if the fluid is Newtonian ($n = 1$), equations (3.18) yield $\alpha_l = \frac{5}{8}$, $A_l = 2.533559\sqrt[4]{\psi}$.

*3.4. Asymptotics for small and large $\xi$. Comparison of exponents.* In the general case, the least of the three numbers $\alpha_k$, $\alpha_\mu$ and $\alpha_l$ defines the asymptotics of the solution of (3.6)-(3.8) when $\xi \to 0$, the largest when $\xi \to \infty$. Therefore, we need to define the mutual positions of these numbers on the real axis.

Since by (3.16) $\alpha_k = \frac{1}{2}$, we have $\alpha_\mu > \alpha_k$ for thinning fluids ($0 < n < 1$) and even for thickening fluids with $1 < n < 2$. For thinning fluids, the number $\alpha_l$, as follows from its definition (3.18), is also greater than $\alpha_k$ in cases of practical significance, for which $\beta_l < 1/n$. For the difference $\alpha_\mu - \alpha_l$, we have:

$$\Delta\alpha = \alpha_\mu - \alpha_l = \frac{n}{2n+2}[\alpha_\mu - (1-\beta_l)]. \tag{3.19}$$

By (3.17) and (3.19), the number $\alpha_l$ is less than $\alpha_\mu$, when $\frac{2}{n+2} > 1 - \beta_l$; it is greater than $\alpha_\mu$, when $\frac{2}{n+2} > 1 - \beta_l$; and the numbers are equal when $\frac{2}{n+2} = 1 - \beta_l$. In the last case, the system (3.6)-(3.8) has the exact monomial solution (3.14) for any $\psi$, when $k = 0$ ($K_{Ic} = 0$), with $\alpha = \alpha_\mu = \alpha_l = \frac{2}{n+2}$ and $A = A_l$ defined by the algebraic equation:

$$A_l^{\frac{n+2}{n}}[(1-\alpha_\mu)B(\alpha_\mu)]^{\frac{1}{n}} = 1 + \frac{\psi}{A_l}.$$

From it, $A_l$ is easily obtained either by plotting the function $\psi(A_l) = A_l^{\frac{2n+2}{n}}[(1-\alpha_\mu)B(\alpha_\mu)]^{\frac{1}{n}} - A_l$, or by proper re-normalizing variables. The exact solution may serve as a benchmark when considering the case $K_{Ic} = 0$ and solving numerically the system (3.6), (3.7).

In HF problems, the difference $\Delta\alpha$ between the viscosity and leak-off exponents is rather small. In the case of a Newtonian fluid ($n = 1$), the maximal difference is $\frac{1}{24}$, when $\frac{1}{3} \le \beta_l \le \frac{1}{2}$. For thinning fluids ($0 < n < 1$), the maximal difference is less than that for a Newtonian fluid and it tends to zero in the limit of perfectly plastic fluid ($n = 0$). The fact that $\alpha_\mu \approx \alpha_l$ may notably complicate numerical solution of the system (3.6)-(3.8), if not accounted for in a numerical scheme. On the other hand, if properly used, it serves to obtain nearly monomial approximate solutions of the types given below.

*3.5. Almost monomial solution for intermediate regimes between toughness dominated and viscosity dominated regimes.* The case of non-negligible toughness while negligible influence of leak-off has been considered in detail for a Newtonian fluid (Garagash and Detournay 2000). The results are summarized in the form of "universal asymptote" as a curve in log-log coordinates in Fig. 2 of the paper by Gordeliy and Peirce (2013). To simplify its employing, we suggest an analytical approximation of the curve in the piece-wise monomial form:

$$\Omega_{k\mu}(\xi_{k\mu}) = A_{k\mu}\xi_{k\mu}^{\,\alpha}, \tag{3.20}$$

where now the opening $w$ and the coordinate $r$ are renormalized as $\Omega_{k\mu} = \frac{w}{w_{k\mu}}$, $\xi_{k\mu} = \frac{r}{L_{k\mu}}$ with $w_{k\mu} = \frac{L_k^2}{L_\mu}$, $L_{k\mu} = \frac{L_k^3}{L_\mu^2}$ and $L_k$ and $L_\mu$ are defined in (3.5). The exponent $\alpha$ and the factor $A_{k\mu}$ are almost constant in wide ranges of the normalized coordinate $\xi_{k\mu}$:

$$\alpha = \begin{cases} 1/2 \\ 0.599, \\ 2/3 \end{cases} A_{k\mu} = \begin{cases} 1 \\ 2^{1/3}3^{5/6} \\ 2^{1/3}3^{5/6} \end{cases} \text{for } \xi_{k\mu} = \begin{cases} \xi_{k\mu} \leq 10^{-5} \\ 10^{-5} \leq \xi_{k\mu} \leq 1. \\ \xi_{k\mu} \geq 1 \end{cases} \quad (3.21)$$

The approximations (3.21) are obtained for a Newtonian fluid. Extension to an arbitrary power-law fluid is as follows. For the exponent $\alpha$, the intermediate number 0.599 is close to the mean of the toughness $\alpha_k = \frac{1}{2}$ and viscosity $\alpha_\mu = \frac{2}{n+2}$ exponents. Therefore, in the general case we may take the intermediate value as $\alpha = \frac{n+6}{4n+8}$. Prescribing $A_{k\mu}$ is even simpler: since the values in the last two lines for $A_{k\mu}$ equal to $A_\mu$ for a Newtonian fluid, we may set them as $A_\mu$, defined in (3.17) in the general case. Then the same piece-wise monomial equation (3.20) with the same normalized variables becomes available for a non-Newtonian fluid.

*3.6. Almost monomial solution for intermediate regimes between viscosity dominated and leak off dominated regimes.* Consider the case important for practice, when the rock toughness may be neglected ($k = 0$). Then the last equation actually drops out of the systems (3.6)-(3.8) and (3.11)-(3.13). In this case, in intermediate calculations, we exclude the parameter $\psi$ by using the renormalized values (3.9), (3.10). Omitting details of the analysis, we present the final result. The almost monomial solution in terms of the formerly used variables $\Omega$ and $\xi$ is:

$$\Omega(\xi) = A_{\mu l}\xi^\alpha, \quad (3.22)$$

where $\alpha$ and $A_{\mu l}$ are piece-wise constant functions of the renormalized coordinate $\xi_1$:

$$\alpha = \begin{cases} \alpha_\mu \\ \alpha_m, \\ \alpha_l \end{cases} A_{\mu l} = \begin{cases} A_\mu(1 + \Delta_\mu \xi_1^{-\varepsilon_\mu}) \\ A_m\psi^{\delta_m} \\ A_l\psi^{\delta_l}(1 + \Delta_l\xi_1^{\varepsilon_l}) \end{cases} \text{for } \xi_1 = \begin{cases} \xi_1 \geq 2 \\ 0.01 \leq \xi_1 \leq 2. \\ \xi_1 \leq 0.01 \end{cases} \quad (3.23)$$

Herein,

$$\xi_1 = \frac{\xi}{\psi^{1/\varepsilon_\mu}}, \ \varepsilon_\mu = \alpha_\mu - (1 - \beta_l), \ \varepsilon_l = \alpha_l - (1 - \beta_l), \delta_l = \frac{\varepsilon_\mu - \varepsilon_l}{\varepsilon_\mu}, \delta_m = \frac{\alpha_\mu - \alpha_m}{\varepsilon_\mu}, \quad (3.24)$$

$$\Delta_\mu = \frac{1}{A_\mu}\frac{n}{n+1+C_{1-\beta_l}/C_\mu}, \Delta_l = A_l\frac{n}{2n+1+C_{\alpha_l+\varepsilon_l}/C_{\alpha_l}}, C_\gamma = (1 - \gamma)B(\gamma),$$

with $B(\gamma)$ given by (3.15), while the parameters $\alpha_m$ and $A_m$ are defined by linear interpolation of the values of $\Omega_1(\xi_1)$ in log-log coordinates between the points $\xi_1 = 0.01$ and $\xi_1 = 2$:

$$\alpha_m = \frac{log\Omega_1(2) - log\Omega_1(0.01)}{2.3010}, \ A_m = \frac{\Omega_1(2)}{2^{\alpha_m}} \quad (3.25)$$

with $\Omega_1(2) = 2^{\alpha_\mu}A_\mu(1 + \Delta_\mu 2^{-\varepsilon_\mu})$ and $\Omega_1(0.01) = 0.01^{\alpha_l}A_l(1 + \Delta_l 0.01^{\varepsilon_l})$.

For certainty, it is assumed that $\alpha_l < \alpha_\mu$. The error of the approximate solution (3.22), (3.23) does not exceed 5% in the entire range of the parameter $\psi$ ($0 \leq \psi < \infty$) and coordinate $\xi$ ($0 \leq \xi < \infty$).

In the particular case of a Newtonian fluid ($n = 1$) and Carter's leak-off ($\beta_l = 1/2$), the definitions of $\alpha_\mu$, $\alpha_l$, $A_\mu$, $A_l$ and equations (3.24), (3.25) yield: $\alpha_\mu = 2/3$, $\alpha_m = 0.6484$, $\alpha_l = 5/8$, $A_\mu = 3.147$, $A_m = 3.634$, $A_l = 2.534$, $\varepsilon_\mu = 1/6$, $\varepsilon_l = 1/8$, $\delta_m = 0.10962$, $\delta_l = 1/4$, $\Delta_\mu = 0.1589$, $\Delta_l = 0.5138$. In this case, the first line in (3.23) actually presents the truncation to two terms for so-called far-field expansion given by Garagash et al. (2011) in equations (3.10), (3.11) of their paper. The third line presents the expansion given by Lenoach (1995) for this case, and it also coincides with the truncated expansion for so-called "intermediate field $\tilde{m}$-expansion", given by Garagash et al. (2011) in equations (3.13),(3.14) of their paper.

For a wide range of the parameter $\psi$, we may compare the universal asymptotics, provided by (3.22), (3.23), with the numerical data of the paper by Lenoach (1995). The author used the following values of parameters: $n = 1$, $\mu = 10^{-7}$ MPa, $E = 2.5 \cdot 10^4$ MPa, $v = 0.15$, $K_{IC} = 0$, $v_* = 0.2$ m/s, $\beta_l = \frac{1}{2}$ and four values of the leak-off factor $C_l$, equal $10^{-5}$, $2.5 \cdot 10^{-5}$, $5.0 \cdot 10^{-5}$ and $7.5 \cdot 10^{-5}$ ms$^{-1/2}$. When using these values, the resulting opening $w$ practically coincides with that given in figures 1-3 of the Lenoach paper for the distance $0 \le r < 5$ from the fracture tip, presented in these figures. Note that the range of the leak-off parameter $C_l$ is wide enough to cover the regimes of low, moderate and large influence of leak-off, described, respectively, by the first, second and third line in (3.23).

## 4. EMPLOYING UNIVERSAL ASYMPTOTICS
## FOR TRACING FRACTURE PROPAGATION

From (3.4) it is clear that the universal asymptotics defines the opening $w = \varphi(d, v_*)$ as a function of the fracture speed $v_*$ at each point of a propagating front with $d$ being the distance from the front. Assume for certainty the dependence monomial:

$$w = A_w(v_*)d^\alpha. \tag{4.1}$$

Inversion of $A(v_*)$ yields the SE in the universal form:

$$v_* = \frac{dx_{*n}}{dt} = A_w^{-1}\left(\frac{w}{d^\alpha}\right). \tag{4.2}$$

For points $\mathbf{x}_i$ of a computational mesh, closest to the front, equation (4.2) gives:

$$\frac{dx_{*ni}}{dt} = A_w^{-1}\left(\frac{w(\mathbf{x}_i)}{d_i^\alpha}\right). \tag{4.3}$$

where $d_i = |\mathbf{x}_{*i} - \mathbf{x}_i|$. Hence after spatial discretization of the global system of equations, we may add the SE (4.3) at nodal points, which are closest to the fracture front and which are under the asymptotic umbrella (we assume that the mesh is not too rough).

In this way, we arrive at a dynamic system. There are numerous options for solving it. In particular, methods of the Runge-Kutta type, employed for 1D problems (Mishuris et al. 2012; Linkov & Mishuris 2013), show high computational potential of such approaches. The conventional LSM and FMM (Sethian 1999) are of use, as well. The last methods look especially promising although they have not been employed for modeling HF so far. The ILSA (Peirce and Detournay 2008) also presents a particular way of using the SE in the form (4.2) based on the universal asymptotics (4.1); in this case, (4.1) is solved in $d$: $d = \left[\frac{w}{A_w(v_*)}\right]^{1/\alpha}$, and $d$ serves as the crossing-time field of the FMM.

*Comment*. In practical calculations, the global mesh is rather rough. Then small difference in the exponents of the monomial approximation (4.1) for viscosity dominated and leak-off dominated regimes does not matter. Commonly, the influence of the difference is beyond the accuracy of calculations. Actually, only the changes in the factor $A_w(v_*)$ in (4.1) are of real significance, while the exponent may be taken fixed.

*Acknowledgement.* The author gratefully acknowledges the support of the FP7 Marie Curie IAPP transfer of knowledge program (project PIAP-GA-2009-251475).